\documentclass[prx,twocolumn,floatfix]{revtex4-2}
\usepackage{latexsym}
\usepackage{graphicx}
\usepackage{subfigure}

\usepackage{subfigure}
\usepackage{array}
\usepackage{verbatim}
\usepackage{amsmath}
\usepackage{color}
\usepackage{xcolor}
\usepackage{soul}
\usepackage{tabu}
\usepackage{multirow}
\usepackage{lipsum}
\usepackage[normalem]{ulem}
\begin{document}
\graphicspath{{fig/}{./}}
%\title{Weyl-type topological  excitonic condensation}
\title{Weyl excitonic condensation}
\author{Efstratios Manousakis}
\affiliation{Department  of  Physics,
  Florida  State  University,  Tallahassee,  Florida  32306-4350,  USA}
\date{\today}
%\section{Introduction}
\begin{abstract}
  We consider a half-filled two-dimensional Su-Schrieffer-Heeger
  lattice and examine the role of the long-range Coulomb electron-hole
  attractive  interaction.
%  and we discover a many-body ground-state characterized
  % by non-trivial topology.
  We demonstrate that, under specific conditions, a rare  interplay of
  topological and  excitonic-collective behavior emerges as a novel
  state of matter.
  %  an unconventional type of
  A unique Bose-Einstein condensate of excitons forms, exhibiting
  co-presence of pseudo-spin
  %(owing to the sublattice-breakup duality) dipole-like
  chiral texture. 
  The  emerging complex order-parameter, a particle-hole pairing-gap, has
  non-zero real and imaginary parts throughout the Brillouin zone (BZ)
  but vanish separately on two different nodal
  lines, which intersect at two Weyl points.
  The Weyl nodes possess opposite pseudo-spin
  chiralities, which act as source and drain of a
  Berry-flux associated with the particle-hole pairing-wavefunction, and are the cause of Bogoliubov-deGennes
  Fermi-arc edge-states.
  We self-consistently calculate the full momentum-dependence of the particle-hole pairing gap
  throughout the entire BZ. Near the Weyl points,
  the pairing gap exhibits
  the unconventional time-reversal-symmetry breaking $p_x+ip_y$ character.
  %which is consistent with the ground-state's pseudo-spin chiral structure.
  Finally, we discuss general potential experimental realizations of
  this novel state of matter.
\iffalse
   We also argue that this novel state of matter can be realized more generally in   
  materials which are in the vicinity of a structural instability,
  such as the one driven by a Peierls or a Jahn-Teller distortion.
  When a material is in such a ground state, it should be expected to exhibit profound effects
  in their transport properties.
  \fi
\end{abstract}
\maketitle
\section{Introduction}
\label{introduction}
Concepts from symmetry and topology as well as the tools
that naturally emerge from their application
are playing a fundamental role in the characterization of electronic structure of materials. In particular,  electronic
topology\cite{PhysRevLett.95.226801,PhysRevLett.95.146802,RevModPhys.83.1057,RevModPhys.82.3045,Bernevig} has recently attracted considerable interest, not only due to the need for a more complete understanding of electronic structure, but also because of its inherent connection to stability, which results in long lifetimes for excitations possessing non-trivial topology.

Among the first ideas to emerge about half a century ago was the Su-Schrieffer-Heeger (SSH)\cite{SSH1979} model of polyacetylene.
 The model, despite its
apparent simplicity, has helped introduce various
novel topological phenomena\cite{Rice1982,Zak1989,SSH_RMP}, including charge fractionalization,
solitons and edge states.
Other important later  developments in the field of electronic topology are
the understanding gained by a detailed analysis of the
fractional quantum Hall-effect\cite{PhysRevLett.49.405,PhysRevB.48.8890},
and the conception and discovery of the topological\cite{PhysRevLett.95.226801,PhysRevLett.95.146802,RevModPhys.83.1057,RevModPhys.82.3045} and the fractional
Chern insulators\cite{PhysRevLett.106.236804,doi:10.1142/S021797921330017X,PhysRevLett.106.236803} and of the
Weyl semimetals\cite{Lu2015,Lv2015,Xu2015,Huang2016,Jiang2017,Armitage2018}.

Another important area to consider for robust, stable, and long-lived behavior is emergent quantum many-body phenomena in solids, like superconductivity\cite{Schrieffer,Phil} and quantum magnetism\cite{Subir,RevModPhys.63.1}, and many other
collective states of matter, which appear as a consequence of macroscopic quantum coherence. Furthermore, the interplay between such collective many-body phenomena
and non-trivial topological character is expected to result in significantly enhanced stability and long coherence-times.
However, with the exception of topological superconductivity\cite{RevModPhys.83.1057,Bernevig} and a few other instances\cite{PhysRevB.108.214431}, topology has primarily been used to characterize single-electron properties.
The goal of the present paper is to provide an interesting example of
a topologically non-trivial behavior of
a collective many-body ground state of an electronic system
characterized by a long-range order.
%with the corresponding order parameter of purely quantum mechanical nature.
On general grounds, such ground-states should be expected to be very stable
and should provide protection from decay of the elementary excitations
above them. Such long-lived excitations should 
find application in areas where long coherence-times are
required\cite{PhysRevB.107.245423,PhysRevLett.86.268,PhysRevX.6.031016,Alicea2012}.

A widely known electrical insulator is the so-called  ``band insulator''
which is a result of single electron behavior. The gap in this insulator
is a mere consequence of the Pauli-exclusion obeying
electron-packing and atomic-level quantization when  these atoms form a weakly interacting periodic array.
A qualitatively different insulating state,  arising from many-body
effects associated with the electron-hole attractive interaction,
is the excitonic insulator\cite{PhysRev.158.462,RevModPhys.40.755,PhysRevLett.74.1633,Eisenstein2004,science.aam6432,Butov_2004,RevModPhys.42.1}. This state arises when such an interaction
is strong enough to cause an instability of the ground-state of the band-insulator  against
electron-hole pairing. In a similar manner to the superconducting
pairing instability of the metallic state, which occurs
in the particle-particle channel\cite{Schrieffer}, the excitonic insulator
is the result of a pairing instability
observed in the particle-hole
channel\cite{PhysRevB.107.075105}. This instability towards an excitonic insulator opens up a pairing gap for electron-hole excitations\cite{PhysRevB.107.075105} and physically
corresponds to a Bose-Einstein condensation (BEC) of electron-hole pairs.
There are several
reports\cite{science.aam6432,Wang2019,PhysRevLett.99.146403,PhysRevB.90.155116,pnas.2010110118}  claiming observation of this form of excitonic condensation
in different materials and under various conditions.
There is also at least one claim that  a topological excitonic insulating state\cite{PhysRevLett.112.176403,PhysRevLett.103.066402} has been
observed\cite{Du2017}.

Since its conception and discovery\cite{Lu2015,Lv2015,Xu2015}  Weyl type-I
semimetallic behavior has been observed
in various compounds\cite{Armitage2018} including type-II
in transition-metal diachalcogenides\cite{Deng2016,Huang2016,Jiang2017,MoTe2Rhodes_PRB2017}. The nature of such behavior is rather
well-understood as a topology that affects the single-electron
properties\cite{Armitage2018}.
In a rather recent paper\cite{PhysRevB.100.104522}
it was demonstrated that if
we stack SSH chains to form a two-dimensional SSH lattice (2D-SSH), the
system, in a specific parameter regime, becomes a Weyl semimetal.
It has also been demonstrated\cite{PhysRevB.100.104522} that
by adding an ad hoc (i.e., without justifying its origin)
attractive interaction between  electron pairs on the same sublattice,
the Weyl semimetal found in the 2D-SSH model immediately becomes
unstable towards a Weyl
superconductor.
We note that there are current experimental attempts\cite{Geng2022}
to make this 2D-SSH lattice; however, it seems that the growth process is
inherently lacking the control required to bring the system in the
appropriate parameter
regime to observe a pure Weyl behavior.
In Sec.~\ref{conclusions} we discuss possible ways to achieve
such a parameter regime.

In the present paper we discuss the instability of the
Weyl semimetallic state of a dual sublattice system, such as the 2D-SSH lattice,
towards an excitonic condensation leading to a condensed state of Weyl topology.
This instability is caused by the {\it realistic} long-range Coulomb interaction
in the particle-hole channel, which leads to formation of
a Bose-Einstein condensate of excitons where the order parameter and the
character of the emerging quasiparticles exhibit all the
signs of a Weyl topology. Furthermore 
near the Weyl points, which appear on the particle-hole pairing order parameter,
the latter is of
the unconventional and highly sought-after time-reversal-symmetry breaking
$p_x+ip_y$\cite{PhysRevB.61.10267}, consistent with the pseudo-spin chiral structure of    the ground-state.

\iffalse
In the present paper, we will use the half-filled SSH lattice
in order to demonstrate the instability
of the Weyl semimetallic state due to a realistic attractive
Coulomb electron-hole interaction towards an excitonic semimetal
where the excitonic gap order parameter exhibits very interesting
topological features.

The model has also been emulated using cold atoms in optical
lattices, where the Zak phase [5] has been measured [7] and
the so-called Thouless charge pumping [8] has been realized
as topological charge pumping [9–12].

 Excitonic insulator as a many-body state and its competition with other forms of insulators, such as the Mott insulator or standard band-insulators or topological or Chern insulator.

5. Here, we propose a new form of many-body ground state with topological order.
\fi
\section{Model}
\label{sec:formulation}
\begin{figure}
    \vskip 0.2 in \begin{center}
            \includegraphics[scale=0.28]{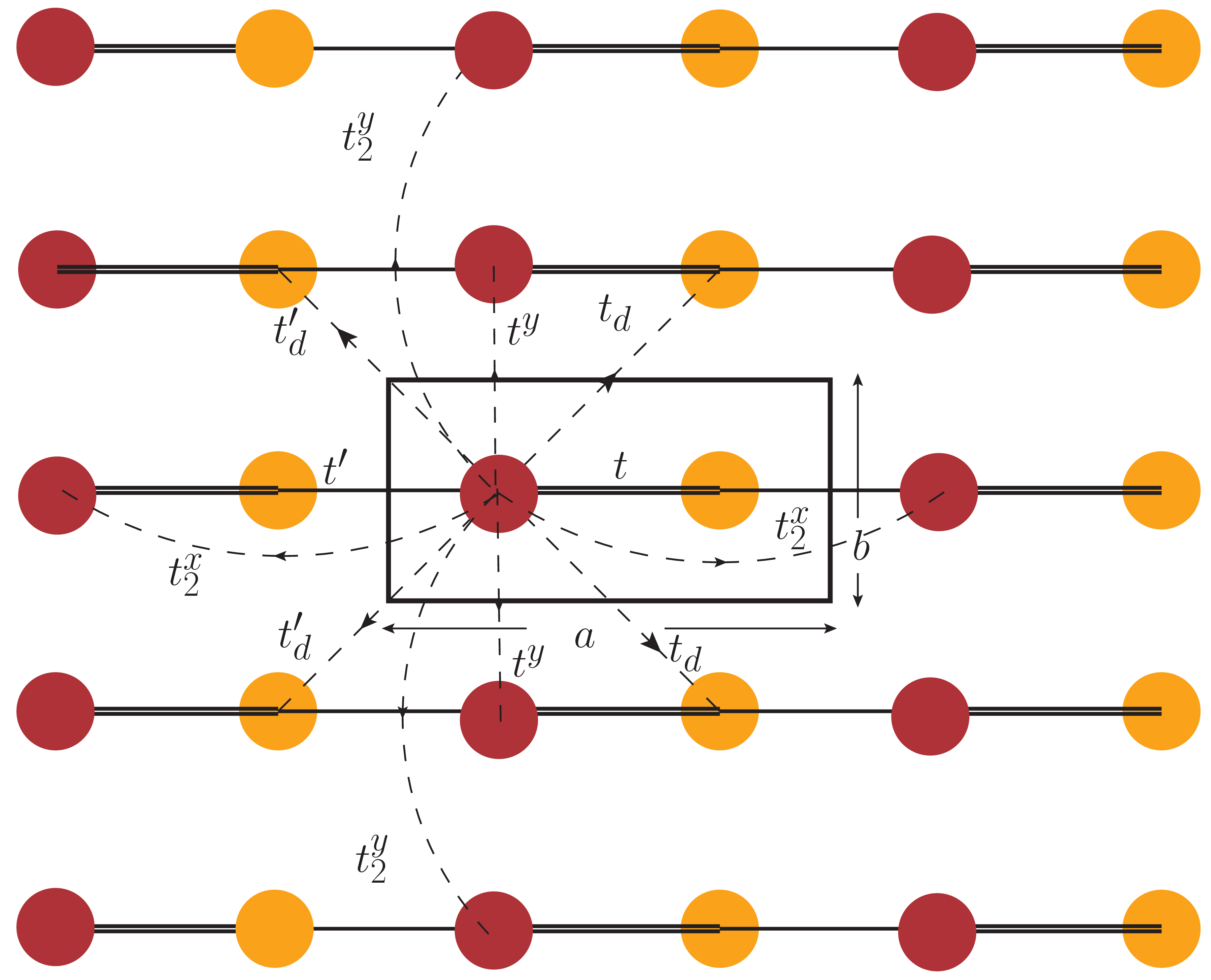} 
     \end{center}
    \caption{Illustration of the SSH lattice. The conventional unit cell is shown as
      the black rectangular with sides $a$ and $b$. The reason for requiring two identical atoms
      in the unit cell is the inversion symmetry breaking because of the
      two different hopping matrix elements $t$ and $t^{\prime}$ caused
      by the dimerization. This leads to the two sublattice ($A$ (garnet) and $B$ (gold)) breakup. }
           \label{fig1} 
\vskip 0.2 in
\end{figure}

We consider the stacking of the well known SSH chains shown in Fig.~\ref{fig1}.
where the unit-cell sizes in the $x$ and $y$ directions are $a$ and $b$.
A simpler version of this system has been studied in Refs\cite{PhysRevB.104.134511,PhysRevB.100.104522,PhysRevB.106.054511} by adding an ad hoc attractive {\it electron-electron}
interactions to investigate the possibility of Weyl superconductivity.
Here, we will investigate the effect of the ubiquitous and realistic {\it electron-hole}
attractive Coulomb interaction.
As illustrated in Fig.~\ref{fig1}, we have considered all possible
nearest neighbor (NN)  and next-nearest neighbor (NNN)
hopping matrix elements. Notice that due to the
dimerization along the $x$-direction, the NN hopping along the
positive $x$-direction is $t$, and along the negative $x$-direction
is $t'$. The NN hoppings along the $\hat y b$ and $-\hat y b$
directions are denoted by $t^{y}$.
The dimerization affects the diagonal hoppings between opposite sublattices, therefore, the two
hoppings along the $\hat x a +\hat y b$ and $\hat x a - \hat y b$ are denoted
as $t_d$, while those  along the  $-\hat x a +\hat y b$ and $-\hat x a -\hat y b$
are denoted by $t^{\prime}_d$. The NNN along the $2 \hat x a$ and
$-2 \hat x a$ are denoted as $t^x_2$, while the NNN along the $2 \hat y b$ and
$-2 \hat y b$ directions are denoted as $t^{y}_2$.
The Hamiltonian of this very tight-binding
model can be written as follows:
\begin{eqnarray}
 \hat {\mathrm{H}}_{\mathrm{SSH}} &=&  \sum_{\vec k}  \Bigl (
           c^{\dagger}_{A}(\vec k),
             c^{\dagger}_{B}(\vec k) \Bigr )
       {\bf M}_0(\vec k)   \begin{pmatrix}
    c_A(\vec k) \\
      c_B(\vec k)  
  \end{pmatrix}, \label{SSH} \\
   {\bf M}_0(\vec k) &=& 
  \begin{pmatrix}
    \epsilon_A({\bf k}) &  -w^*_{\vec k} \\
      -w_{\vec k} & \epsilon_B({\bf k})  
  \end{pmatrix}, \hskip 0.1 in
  w_{\vec k} = w^x_{\vec k} - i w^y_{\vec k},\label{ssh-w}\\
  w^x_{\vec k} &=& {\alpha}_{\vec k} + {\beta}_{\vec k} \cos(k_x a), \hskip 0.1 in
  w^y_{\vec k} =  {\gamma}_{\vec k} \sin(k_x a),   
\end{eqnarray}
where
\begin{eqnarray}
{\alpha}_{\vec k} &=& t + 2 t_d \cos(k_y b),\\
{\beta}_{\vec k} &=& t^{\prime} + 2 t^{\prime}_d \cos(k_y b),\\
{\gamma}_{\vec k} &=& t^{\prime} + 2 t_d \cos(k_y b).
\end{eqnarray}
The diagonal matrix elements $\epsilon_A({\bf k})$ and $\epsilon_B({\bf k})$
are the energy dispersion obtained from the on-site energies
$\epsilon^0_{A,B}$ and the hoppings $t^{y}({\nu})$, $t^x_{2}(\nu)$ and $t^{y}_{2}(\nu)$ (for the general case, the added index $\nu =A,B$
stands or sublattices A and B)
between the same sublattice (depicted in Fig.~\ref{fig1}). They are given as
\begin{eqnarray}
  \epsilon_{\nu}({\bf k}) &=& \epsilon^0_{\nu} - 2 t^y({\nu}) \cos(k_y b) - 2 t^x_{2}(\nu) \cos(2k_x a) \nonumber \\
  &-&2 t^{y}_{2}(\nu) \cos(2 k_y b), \label{AtoA}
\end{eqnarray}
where $\nu=A,B$. In general the onsite energies and the hopping
for the two sublattice can be different,
however, in our case, as shown below, $\epsilon_A({\bf k}) = \epsilon_B({\bf k})$,
therefore, we will take $t^{y}({\nu})=t_{y}$, $t^x_{2}({\nu})=t^x_2$ and $t^y_{2}({\nu})=t^y_2$
and we will use as origin of the energy axis the value
of the sublattice independnent on-site energy $\epsilon^0_\nu$.

The above matrix $  {\bf M}({\bf k})$ can be
also written in terms of the Pauli matrices as follows
\begin{eqnarray}
  {\bf M}(\bf k) &=& {\bar \epsilon({\bf k})} I -{\vec h}({\vec k}) \cdot \vec \sigma \\
  {\vec h}({\bf k}) &=& \begin{pmatrix}
    w^x_{\vec k} \\
    w^y_{\vec k} \\
    \delta \epsilon({\bf k}) \label{hvector}\\
  \end{pmatrix},
\end{eqnarray}
where
\begin{eqnarray}
\bar \epsilon({\bf k}) = {{\epsilon_A({\bf k}) + \epsilon_B({\bf k})} \over 2}\hskip 0.1 in
\delta \epsilon({\bf k}) = {{\epsilon_A({\bf k}) - \epsilon_B({\bf k})} \over 2},
\end{eqnarray}
and $I$ is the $2\times 2$
identity matrix.
This expression justifies the commonly used term ``pseudo-spin'' when we refer to states on
sublattice A (as pseudo-spin up) or B (as pseudo-spin down).

The atoms on A and B sublattice are the same but there is inversion symmetry breaking due to the dimerization along the $x$-direction. Therefore, if we move
from a $C$-atom on the A sublattice to a $C$-atom on the 
B sublattice and we also transform $x \to -x$ we see the same crystal-environment, which implies that 
\begin{eqnarray}
  \epsilon_A(k_x,k_y) = \epsilon_B(-k_x,k_y).\label{c1}
  \end{eqnarray} 
In addition, because of the inversion symmetry along the $y$-axis, the bands are
invariant under $k_y \to -k_y$, i.e.,
\begin{eqnarray}
  \epsilon_{\nu}(-k_x,k_y) = \epsilon_{\nu}(-k_x,-k_y).
  \end{eqnarray}
where $\nu =A,B$.
Furthermore, due to the time-reversal symmetry (we have no
spin in our problem) we must have
\begin{eqnarray}
\epsilon_{\nu}(k_x,k_y) =
\epsilon_{\nu}(-k_x,-k_y)
\end{eqnarray}
Therefore,
\begin{eqnarray}
\epsilon_B(-k_x,k_y) = \epsilon_B(-k_x,-k_y)
= \epsilon_B(k_x,k_y).
\end{eqnarray}
Combining the above result and Eq.~\ref{c1}
we conclude that $\epsilon_A(k_x,k_y) = \epsilon_B(k_x,k_y)$
Therefore, in the following we will consider the case where $\epsilon_A({\bf k}) = \epsilon_B({\bf k})$.
This implies that there is no $\sigma_z$ term, i.e., the
vector $\vec h({\bf k})$ in Eq.~\ref{hvector} has no $z$ component.

  The Weyl nodes are found by solving the following two equations
\begin{equation}
(t^{\prime} + 2 t^{\prime}_d \cos(k_y b))\cos(k_xa)
  + t + 2 t_d \cos(k_y b) =0,
  \end{equation}
\begin{equation}
(t^{\prime} + 2 t_d \cos(k_y b)) \sin(k_x a) = 0,
\end{equation}
  with respect to $k_x$ and $k_y$.
  There are two or four Weyl nodes, depending of the difference between
  of diagonal hoppings $t_d$ and $t'_d$. Both pairs are obtained
  when $\sin(k_x a) =0$. The first pair is obtained when $k_x=0$:
  \begin{eqnarray}
    k_x&=&0, \\
    k_y &=& \cos^{-1}((t+t^{\prime})/2(t_d+t^{\prime}_d)),
  \end{eqnarray}
   Notice that, in order to have these Weyl nodes, the hoppings $t,t^{\prime},t_d, t^{\prime}_d$ should satisfy the condition:
   \begin{eqnarray}
     t + t^{\prime} \le 2 (t_d + t^{\prime}_d),
   \end{eqnarray}
   while all other hoppings, namely, $t^{''}$, $t^x_2$ and $t^{'''}$ can take
   any values as they only influence the topologically trivial part of the
   Hamiltonian.

\begin{figure*}
    \vskip 0.2 in \begin{center}
            \includegraphics[scale=0.35]{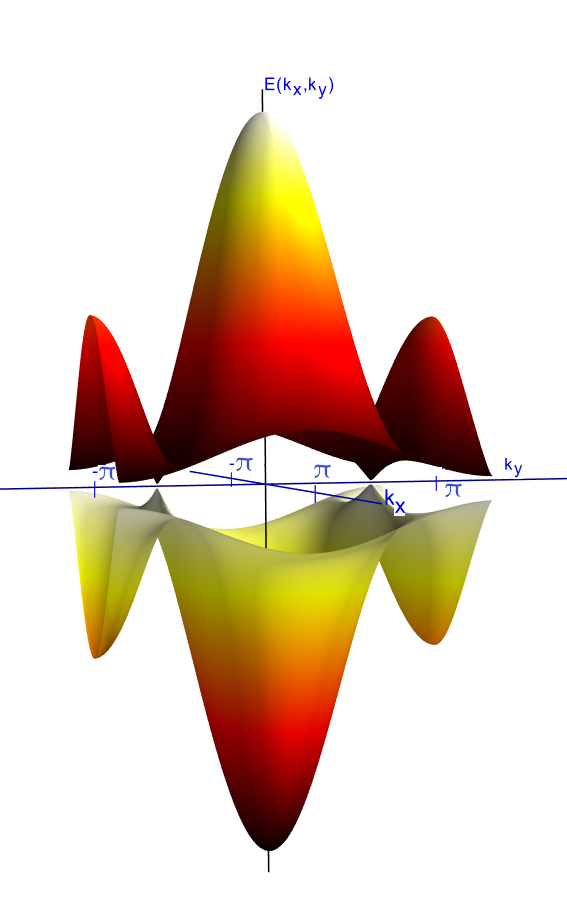} \label{fig2a} \hskip 0.2 in
            \includegraphics[scale=0.35]{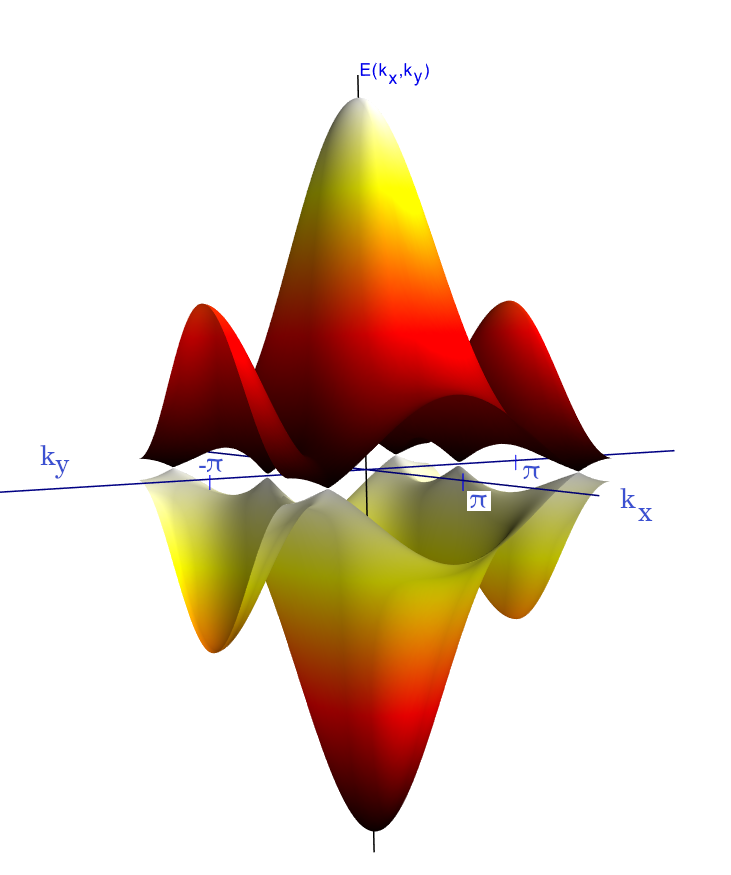} \label{fig2b}
     \end{center}
    \caption{(Left) The non-interacting bands of the Hamiltonian given by Eq.~\ref{SSH}
      for $t=0.6$ $t^{\prime} = 1.2$ and $t_d =1.1$ and $t'_d=1.3$.
      Note the display of
      the two Dirac-Weyl opposite chirality nodes along the $k_y$ axis.
    (Right) The bands when we choose
       $t=0.6$ $t^{\prime} = 1.2$ and $t_d =1.0$ and $t'_d=1.4$.}
           \label{fig2} 
\vskip 0.2 in
\end{figure*}

\begin{figure}
    \vskip 0.2 in \begin{center}
      \includegraphics[scale=0.28]{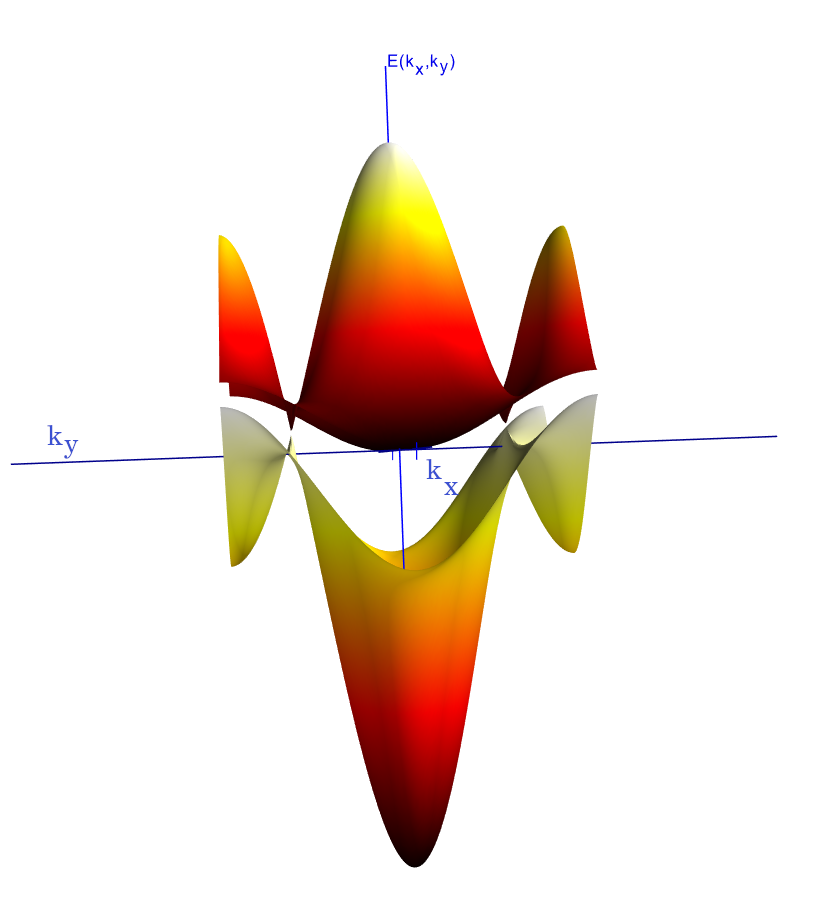}
           \end{center}
    \caption{The non-interacting bands of the Hamiltonian given by Eq.~\ref{SSH}
      for $t=0.6$ $t^{\prime} = 1.2$ and $t_d =1.1$ and $t'_d=1.3$ and $t^{y}=0.5$.}
           \label{fig3} 
\vskip 0.2 in
\end{figure}

   The other pair of Weyl nodes is obtained when $k_x a =\pi$
   and it is given by
  \begin{eqnarray}
    k_x&=&\pm \pi, \\
    k_y &=& \cos^{-1}((t-t^{\prime})/2(t_d-t^{\prime}_d)),
  \end{eqnarray}
  For this pair to be a solution, we must have
    \begin{eqnarray}
      |t - t^{\prime}| \le 2 |t_d - t^{\prime}_d|.
    \end{eqnarray}
    However, we expect that this condition is not likely to be realistically fulfilled because the effect of the dimerization should be affecting significantly more the
    asymmetry $\delta t \equiv |t - t'|$ between the direct bonds involved in the dimerization than the asymmetry $\delta t_d =|t_d-t'_d|$ between the NNN hoppings.
    So, in the rest of the paper we will assume that
    $|\delta t_d| < 2 |\delta t|$.
      The two cases are illustrated in Fig.~\ref{fig2}.
In order to illustrate the role of the so-called topologically trivial
terms from the hoppings from $A \to A$ and $B \to B$, in Fig.~\ref{fig3}
we allow $t^{y}$ (the nearest neighbor interchain hopping) to be non-zero.
The result of keeping these topologically trivial terms is simply to add
to the Weyl part the contribution of the dispersion $\epsilon({\bf k})$.
This contribution simply raises the two Weyl
       points but it does not affect their position, nor their character.
In the following, we will ignore the contribution of the identity matrix
(i.e., of these terms entirely) as it is a topologically trivial contribution.

\begin{figure}
    \vskip 0.2 in \begin{center}
            \includegraphics[scale=0.32]{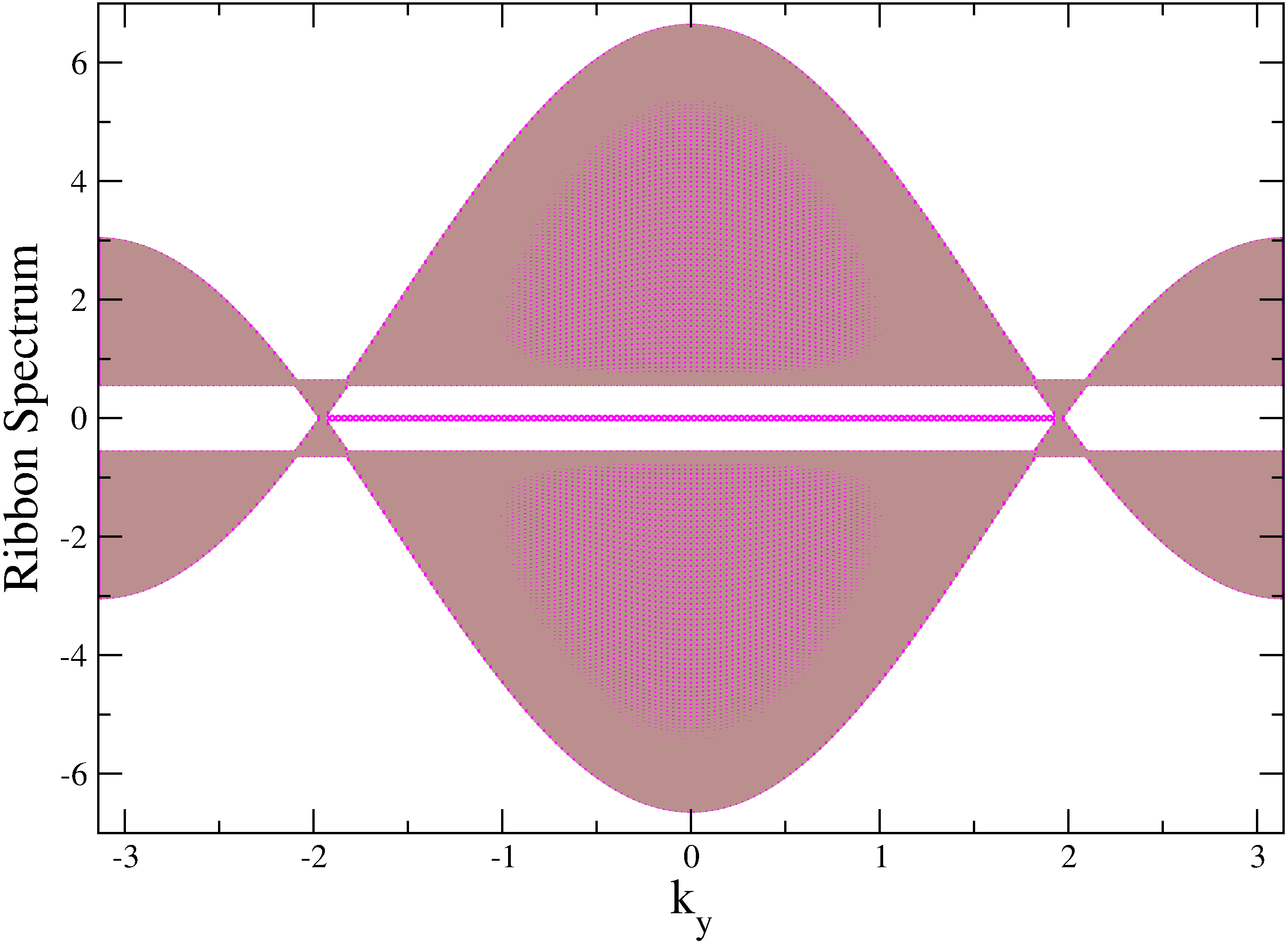} 
     \end{center}
    \caption{The states of a ribbon. The purple dispersionless
      bands   represent two
      Fermi-arc states which are localized on the opposite
      ribbon edges.}
           \label{fig4} 
\vskip 0.2 in
\end{figure}
The two bands that can be found by a straightforward diagonalization
are given as $E_{\pm}(\vec k) = \pm \sqrt {|w^x(\vec k)|^2+|w^y({\vec k})|^2}$
  and are shown in Fig.~\ref{fig2}. The corresponding eigenstates
  are created by the following operators
  \begin{eqnarray}
    \chi^{\dagger}_{\pm} &=& {1 \over {\sqrt{2}}} (c^{\dagger}_A(\vec k)
    \mp e^{-i \phi(\vec k)}  c^{\dagger}_B(\vec k)),\\
    \phi(\vec k) &=& \tan^{-1}\Bigl ({{w^y(\vec k)} \over {w^x(\vec k)}}\Bigr ).
  \end{eqnarray}
  and they also carry  opposite pseudo-spin helicity.

    The which act as source and sink of Berry flux.
   When we carry out a diagonalization
   of a ribbon which is infinite along the $k_y$ direction and finite in the
   $k_x$ direction, we obtain the spectrum of states illustrated
   in Fig.~\ref{fig4}. On the same graph, we have also projected
   the energy of all
   bulk states as a function of $k_y$ for all values of $k_x$.
   Notice that with the exception of two Fermi-arc states (which are bound
   on each of the ribbon edges) shown by the dispersionless
   lines connecting the Weyl nodes, the band of the other states
   exactly overlap with the energy of the bulk states.

Next, we will include the interaction between electrons excited in
band $A$ and holes in band $B$  as in Ref.~\cite{PhysRevB.107.075105}, i.e., 
\iffalse
\begin{eqnarray}
 && \hat V = {1 \over {2}}\sum_{\vec k,\vec k^{\prime},\vec q}
  V_{\vec k\vec k^{\prime}}(\vec q)     c^{\dagger}_{A}(\vec k+\vec q) c^{\dagger}_{B}(\vec k^{\prime}-\vec q)
  c_{B}(\vec k^{\prime}) c_{A}(\vec k),
      \label{inter-exciton}
\end{eqnarray}
\fi
%We can define as a reference state, the ground-state $|\Psi_0 \rangle$ of
%the conventional insulator.
by means of the following form:
\begin{eqnarray}
  \hat V = -{1 \over {2} } \sum_{\vec k,\vec k^{\prime},\vec q}
  V_{\vec k\vec k^{\prime}}(\vec q)    h^{\dagger}_{B}(\vec k+\vec q)  c^{\dagger}_{A}(\vec k^{\prime}-\vec q)  c_{A}(\vec k^{\prime}) h_{B}(\vec k),
  \label{inter-exciton4}
\end{eqnarray}
where we have used the hole creation and annihilation operators
\begin{eqnarray}
  h^{\dagger}_{B}(-\vec k) \equiv c_{B}(\vec k), \hskip 0.2 in 
  h_{B}(-\vec k) \equiv c^{\dagger}_{B}(\vec k).
\end{eqnarray}
Notice that this transformation leads to an attractive interaction between
electrons  and holes which
can lead to formation of excitons which can form a BEC.
In the weak coupling limit, this
phenomenon can be approached as a BCS-like pairing\cite{PhysRev.158.462,RevModPhys.40.755}. Our approach\cite{PhysRevB.107.075105} is to apply the Bogoliubov-Valatin factorization:
\iffalse
as follows:
\begin{eqnarray}
  && h^{\dagger}_{B}(\vec k+\vec q)  c^{\dagger}_{A}(\vec k^{\prime}-\vec q)  c_{A}(\vec k^{\prime}) h_{B}(\vec k) \to \nonumber \\
 && \langle h^{\dagger}_{B}(\vec k+\vec q)  c^{\dagger}_{A}(\vec k^{\prime}-\vec q) \rangle  c_{A}(\vec k^{\prime}) h_{B}(\vec k) \nonumber \\
  &+&
  h^{\dagger}_{B}(\vec k+\vec q)  c^{\dagger}_{A}(\vec k^{\prime}-\vec q) \langle  c_{A}(\vec k^{\prime}) h_{B}(\vec k) \rangle,
\end{eqnarray}
leads to the following approximation
\fi
\begin{eqnarray}
  \hat V_{BV} & = &  -  \sum_{\vec k}
  \Bigl ( \Delta (\vec k)  c_{A}(\vec k) h_{B}(-\vec k)
  + h.c. \Bigr ), \\
  \Delta(\vec k)  &\equiv&
  {1 \over {2} }\sum_{\vec q}   V_{\vec k,-\vec k}(\vec q) \lambda^*(\vec k + \vec q),
  \label{inter-exciton-linearized}\\
  \lambda^*(\vec k) &\equiv& \langle h^{\dagger}_{B}(-\vec k)  c^{\dagger}_{A}(\vec k) \rangle,
  \label{eq:lambda}
\end{eqnarray}
which, apart from a constant, leads to the Bogoliubov-deGennes Hamiltonian:
\begin{eqnarray}
  \hat {\mathrm{H}}_{\mathrm{BdG}} &=& \hat {\mathrm{H}}_{\mathrm{SSH}} - \sum_{\vec k}
  \Bigl ( \Delta (\vec k)  c_{A}(\vec k) h_{B}(-\vec k)
  + h.c. \Bigr ).
\end{eqnarray}
In the case where the self-consistently determined value of $\Delta(\vec k)$ is non-zero, there would be pairing of the electron-hole system.
Both terms can be combined to obtain (apart from a constant) the following form
\begin{eqnarray}
 &&\hat {\mathrm{H}}_{\mathrm{BdG}} =  {1 \over 2 } \sum_{\vec k} {\bf c}^{\dagger}({\vec k})
       {\bf M}(\vec k) {\bf c}({\vec k}),\\
   &&    {\bf c}^{\dagger}({\vec k}) \equiv  \Bigl (
           c^{\dagger}_{A}(\vec k),
             h^{\dagger}_{B}(-\vec k), 
             c_{A}(\vec k),
             h_{B}(-\vec k) \Bigr ),
\end{eqnarray}
where
\iffalse
        &&  C = { 1 \over 2} (\epsilon_{A}
       - \epsilon_{B} + \Delta(\vec k)
       + \Delta^*(\vec k)).
       \fi
\begin{eqnarray}
  {\bf M}(\vec k) &\equiv& 
  \begin{pmatrix}
    \epsilon_{A}({\bf k}) & 0 & 0 &  z^*(\vec k) \\
    0 & -\epsilon_{B}({\bf k}) &  -z^*(\vec k) & 0 \\
    0 &  -z(\vec k) & -\epsilon_{A}({\bf k}) & 0  \\
      z(\vec k) & 0 & 0 & \epsilon_{B}({\bf k})
  \end{pmatrix}, \\
  z(\vec k) &\equiv& \Delta(\vec k) - w(\vec k). \label{zeta}
\end{eqnarray}

After diagonalization, the Hamiltonian (apart from a constant) takes the form:
\begin{eqnarray}
  \hat {\mathrm{H}}_{\mathrm{BdG}} &=& \sum_{\vec k,\nu=\pm} {\cal E}_{\nu}(\vec k)
  \gamma^{\dagger}_{\nu}(\vec k)\gamma_{\nu}(\vec k),
\end{eqnarray}
where the quasiparticle energy bands and corresponding
quasiparticle creation operators are given as follows
\begin{eqnarray}
  {\cal E}_{\pm}(\vec k) &=& \bar \epsilon({\bf k}) \pm \sqrt {(\delta
    \epsilon({\bf k}))^2
    + | z(\vec k)|^2},\\
\gamma^{\dagger}_{\pm}(\vec k) &=& u_{\pm}(\vec k) c_A(\vec k) + v_{\pm}(\vec k)
h^{\dagger}_B(-\vec k)\label{quasiparticle},
\end{eqnarray}
and the coefficients (which depend on an arbitrary overall phase)
can be chosen as 
\begin{eqnarray}
u_{\pm}(\vec k) &=& {{|z(\vec k)|} \over {\sqrt{|z(\vec k)|^2 + ({\cal E}_{\pm}(\vec k) - \epsilon_A({\bf k}))^2}}},\\
v_{\pm}(\vec k) &=& {{{\cal E}_{\pm}(\vec k) - \epsilon_A({\bf k})} \over {\sqrt{|z(\vec k)|^2 + ({\cal E}_{\pm}(\vec k) - \epsilon_A({\bf k}))^2}}}{{z(\vec k)} \over {|z(\vec k)|}}. \label{qp-wave}
\end{eqnarray}
The ground state of the system with a number of electrons just enough
to fill the lowest band is defined as
\begin{eqnarray}
  \gamma^{\dagger}_-(\vec k) | \Psi_0 \rangle = 0, \hskip 0.2 in
  \gamma_+(\vec k) | \Psi_0 \rangle = 0, \hskip 0.2 in \forall \hskip 0.1 in \vec k. \label{ground}
  \end{eqnarray}

\begin{figure*}
    \vskip 0.3 in \begin{center}
        \subfigure[]{
            \includegraphics[scale=0.16]{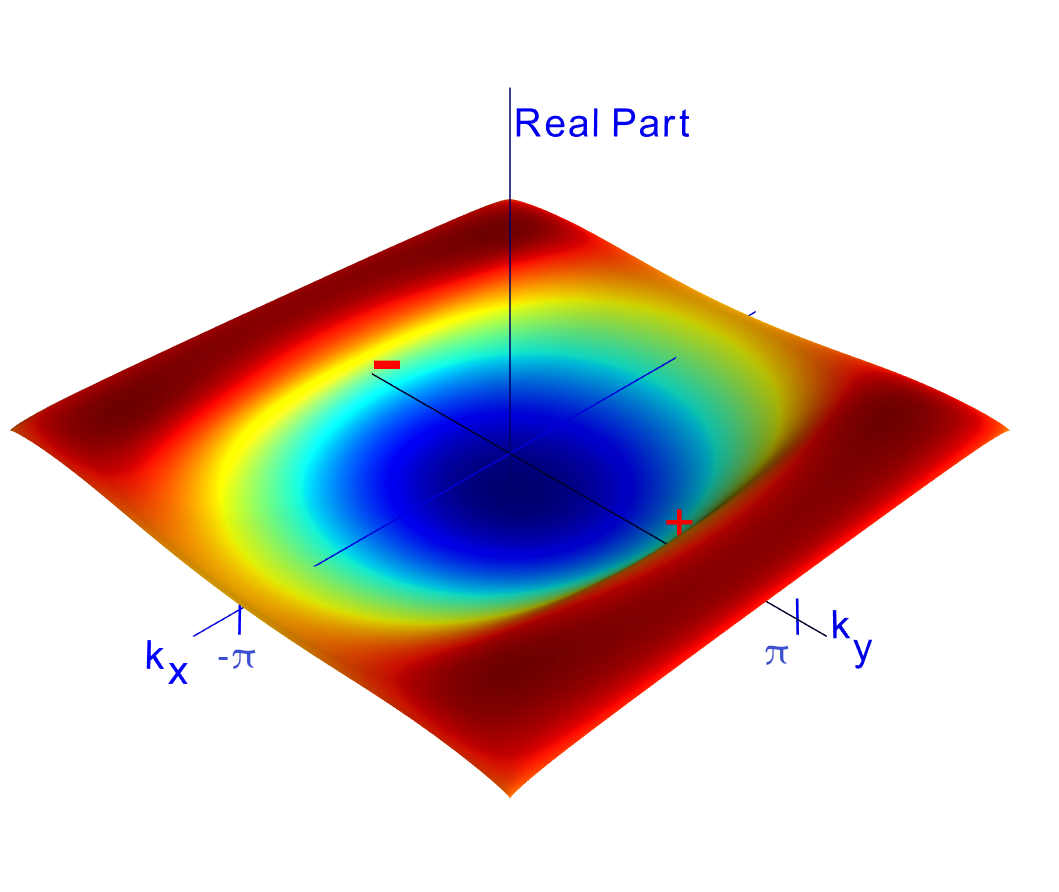} 
        } 
        \subfigure[]{
            \includegraphics[scale=0.14]{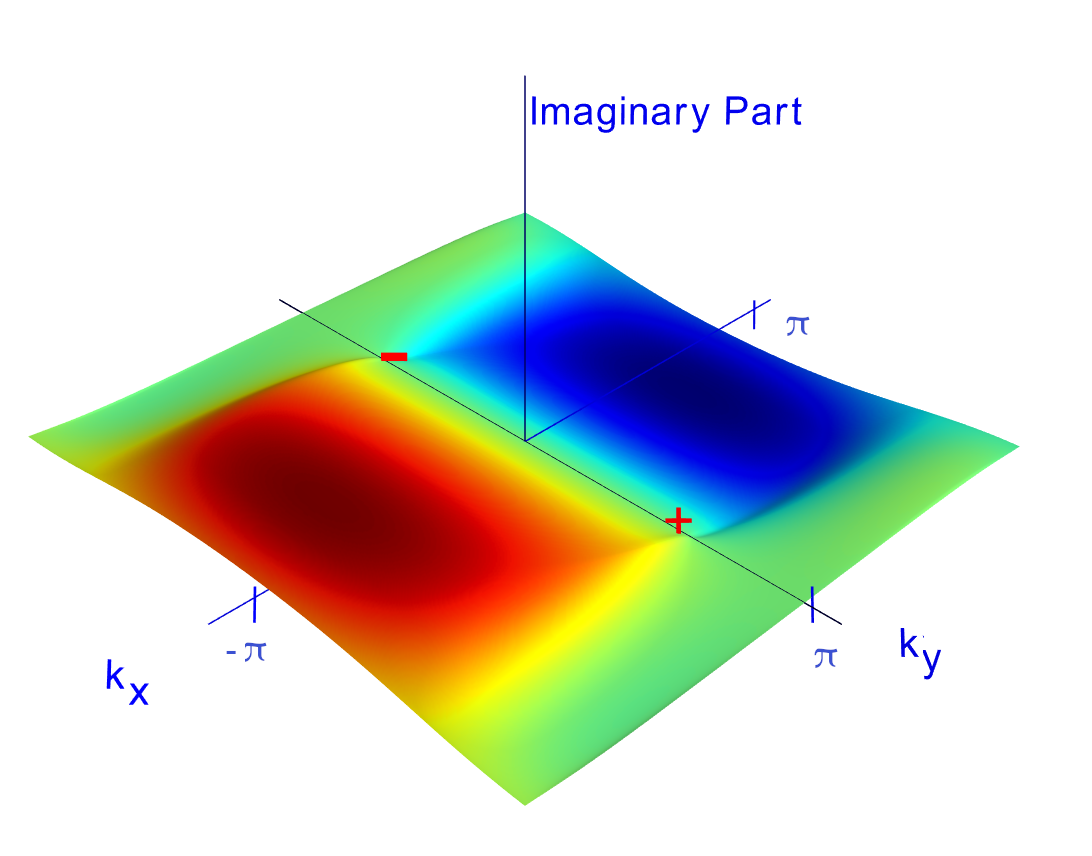}
        }
    \caption{Real (panel (a))  and imaginary part (panel (b)) of the gap $\Delta(k_x,k_y)$
      obtained by solving Eq.~\ref{eq:gap} iteratively until self-consistency was achieved.}
           \label{fig5} 
     \end{center}
\vskip 0.2 in
\end{figure*}

\begin{figure}
    \vskip 0.2 in \begin{center}
            \includegraphics[scale=0.25]{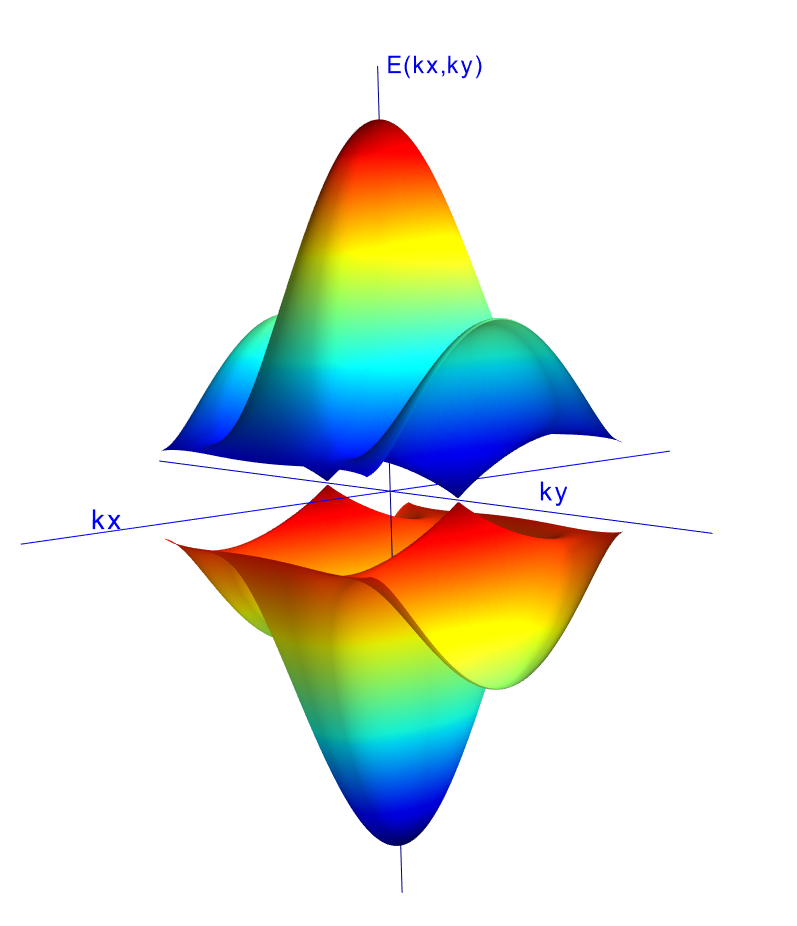} 
     \end{center}
\caption{The bands obtained by solving Eq.~\ref{eq:gap} iteratively until self-consistency was achieved. }
           \label{fig6} 
\vskip 0.2 in
\end{figure}

\begin{figure}
%    \vskip 0.2 in \begin{center}
        \subfigure[]{
            \includegraphics[scale=0.2]{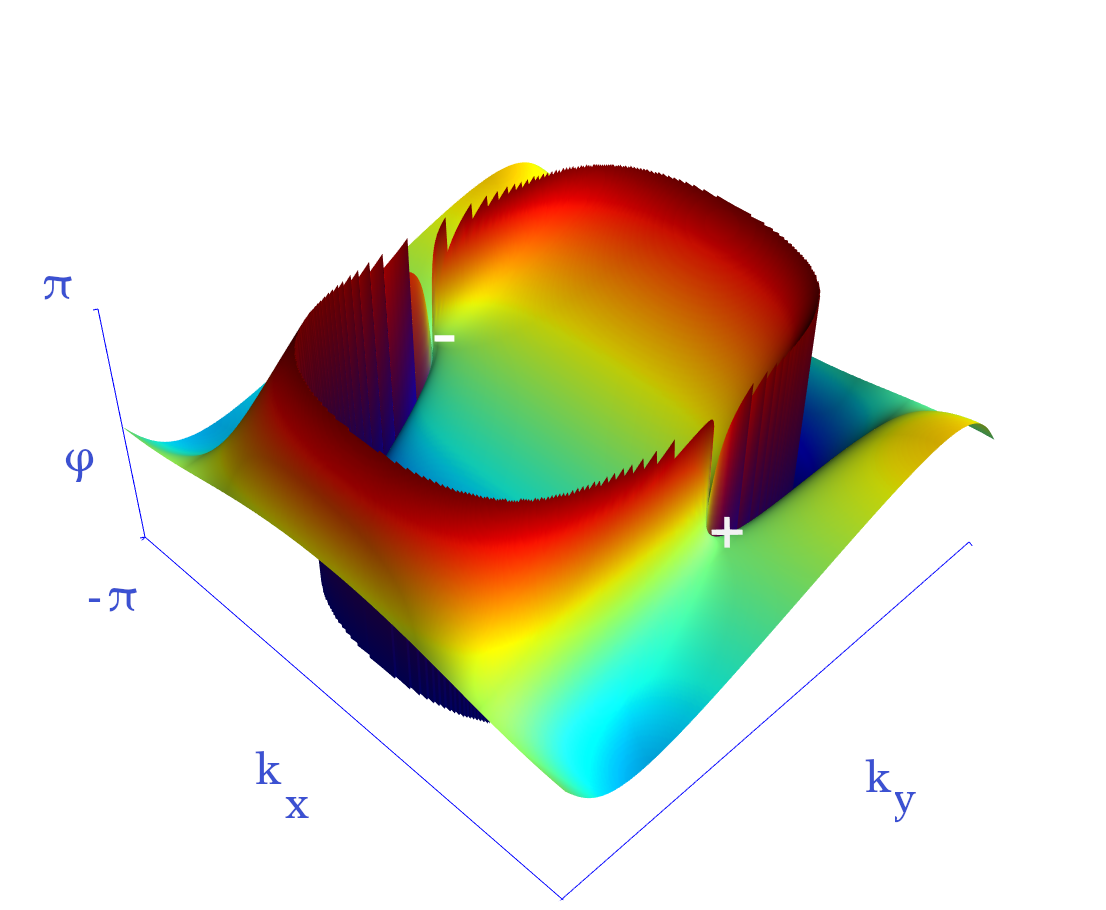} 
        } 
        \caption{The phase of the pairing wave-function given by Eq.~\ref{phase}.
          The two points where the branch cut starts and ends are shown
      as + and -.
      They are the beginning and the end of the branch cut via which you pass from the $-\pi/2$ (blue-colored)  to the $+\pi/2$ (red-colored) Riemann sheet.}
%     \end{center}
            \label{fig7}
\vskip 0.2 in
\end{figure}

\begin{figure}
  \vskip 0.2 in
          \subfigure[]{
            \includegraphics[scale=0.6]{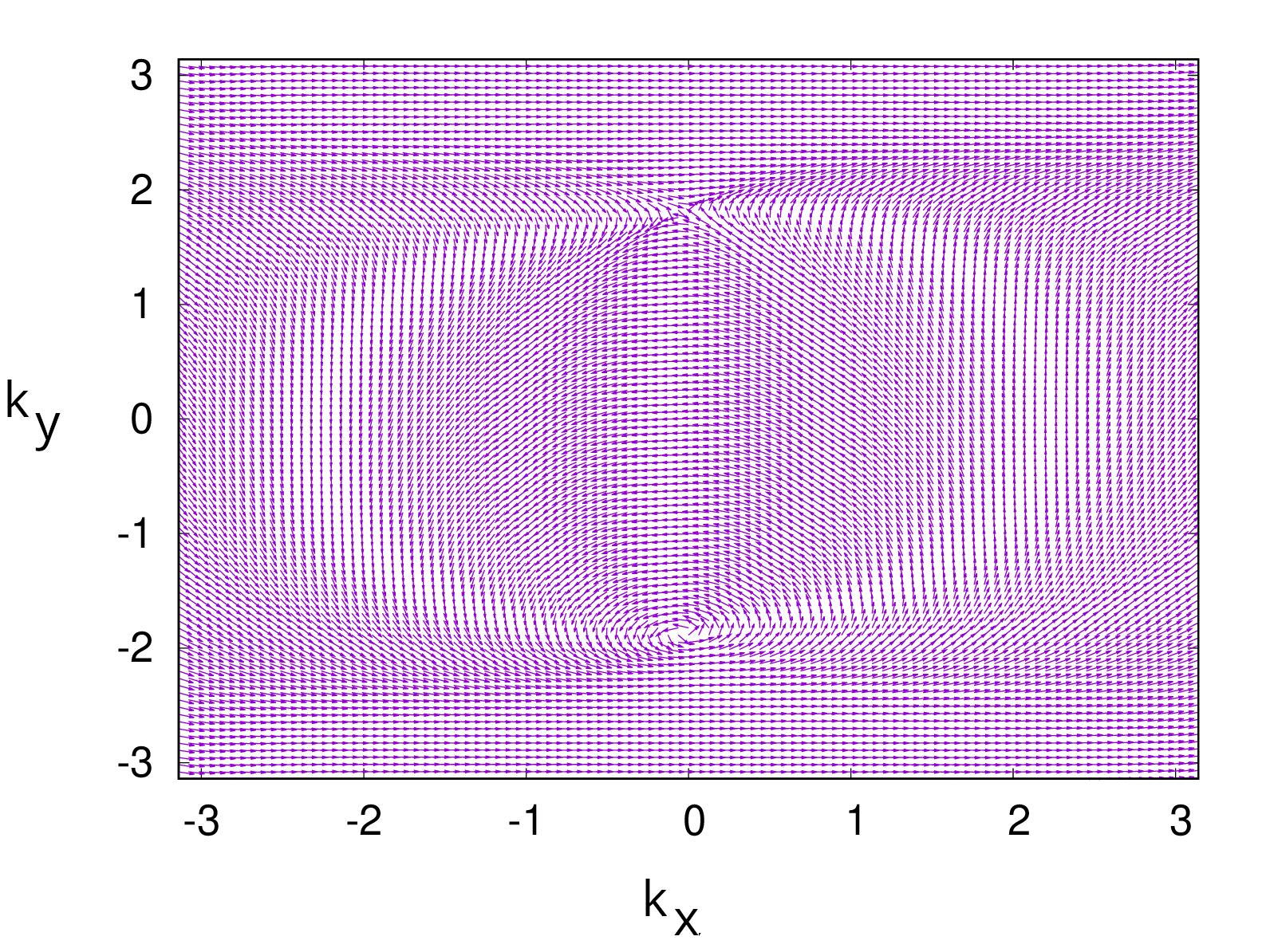} 
        } \\ 
        \subfigure[]{
            \includegraphics[scale=0.3]{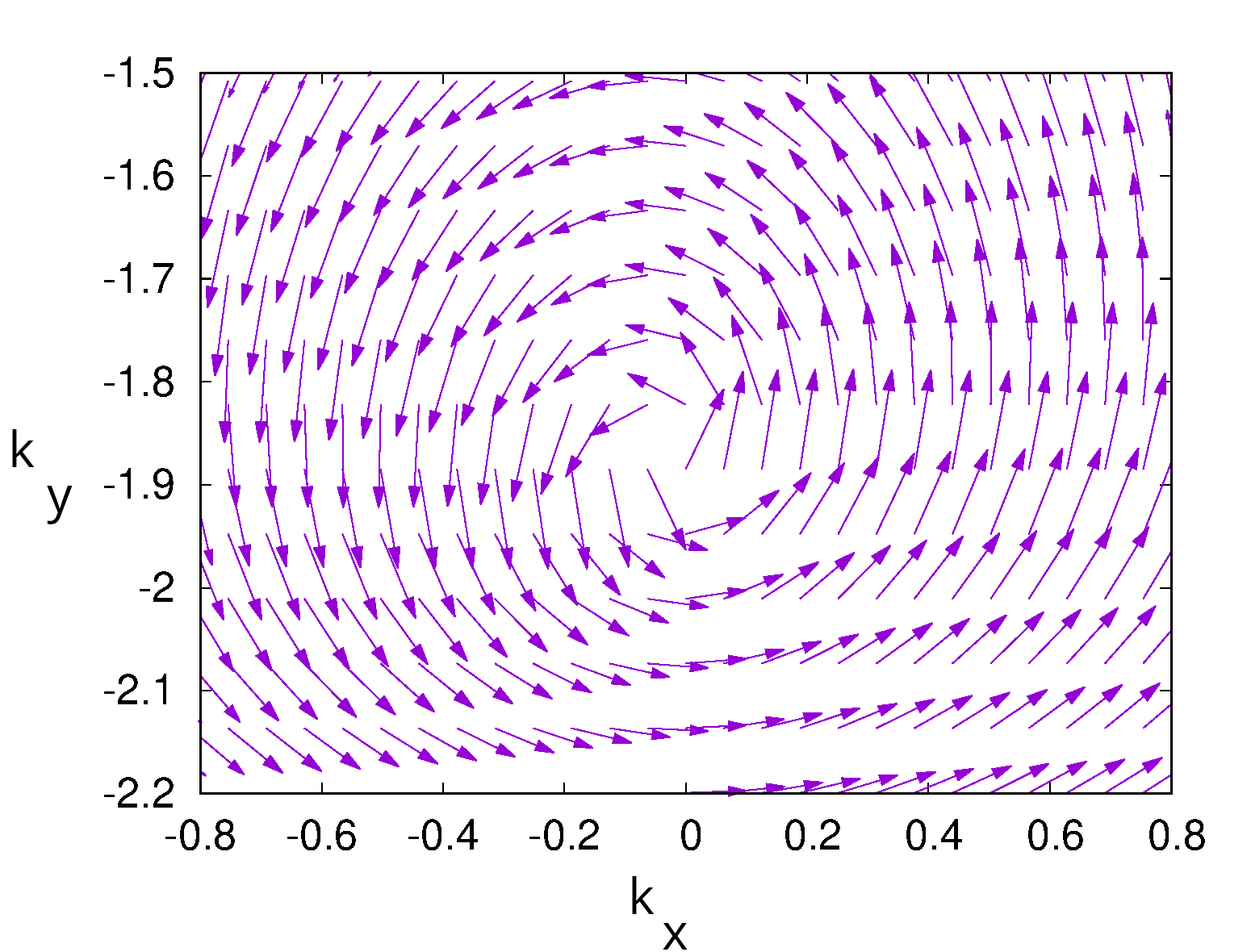}
        }
        \subfigure[]{
            \includegraphics[scale=0.3]{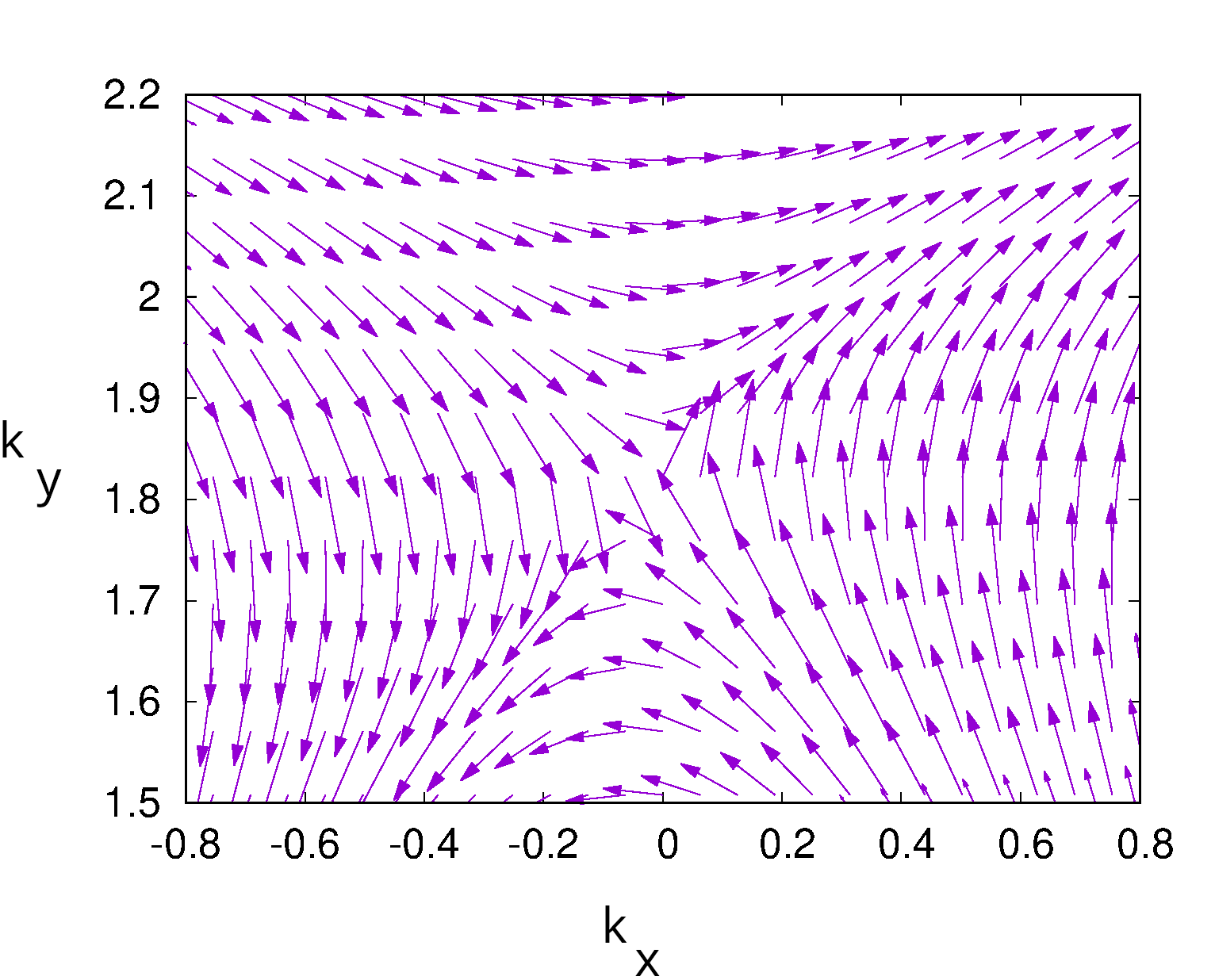}
        }
        \caption{(a) The vector which represents the direction of the order parameter $\vec B({\bf k})$. 
      Both BZ axes $k_x$ and $k_y$  extends from $-\pi$ to $\pi$, i.e., we have taken the unit cell dimensions $a=1$
and $b=1$. For the purpose of illustration we have normalized the vector $\vec B$. Panels (b) and (c) are parts of panel (a) focused around the two Weyl nodes.}
            \label{fig8}
\vskip 0.2 in
\end{figure}

\begin{figure}
    \vskip 0.3 in \begin{center}
            \includegraphics[scale=0.35]{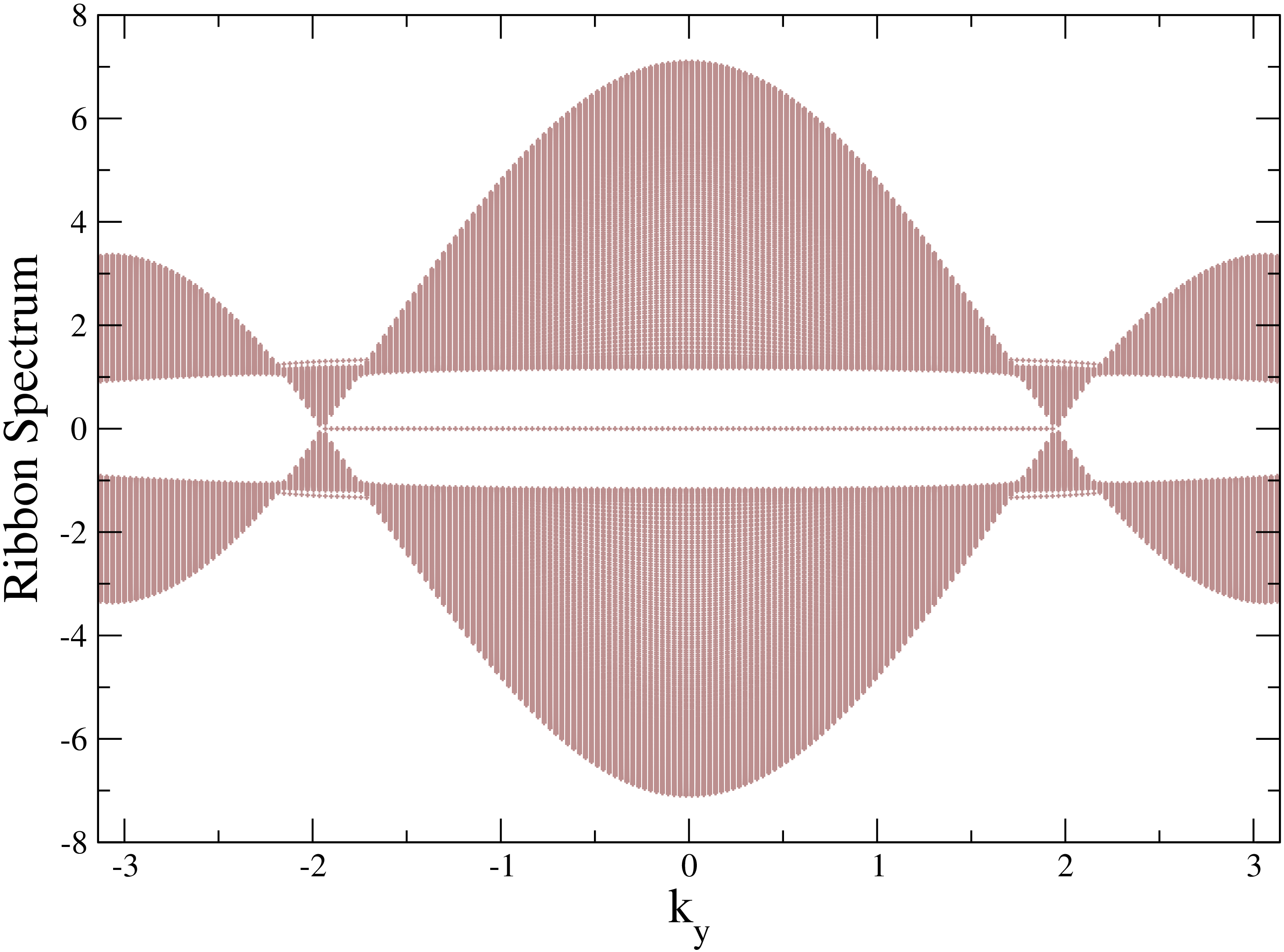} 
     \end{center}
    \caption{States of a ribbon infinite in the $y$ direction and finite in the
      $x$ direction. There is a Bogoliubov Fermi-arc state connecting the
      bulk Weyl points. Notice that the space between the upper and lower
    band is wider than in the non-interacting case of Fig.~\ref{fig4}.}
           \label{fig9} 
\vskip 0.2 in
\end{figure}

By inverting the relations given by Eq.~\ref{quasiparticle} to express $c_A(\vec k)$, $h_B(-\vec k)$, $c^{\dagger}_A(\vec k)$, $h^{\dagger}_B(-\vec k)$
in terms of $\gamma_{\pm}(\vec k)$, $\gamma^{\dagger}_{\pm}(\vec k)$, we can calculate
$\lambda$ given by Eq.~\ref{eq:lambda}, i.e., using the expression
\begin{eqnarray}
  \lambda(\vec k) &=& \langle \Psi_0|  c_{A}(\vec k) h_{B}(-\vec k) | \Psi_0 \rangle,
\end{eqnarray}
and Eqs.~\ref{ground}. It is straightforward to find that:
\begin{eqnarray}
  \lambda(\vec k) &=&
         {{z^*(\vec k)} \over {2
           \sqrt{|z(\vec k)|^2+(\delta \epsilon({\bf k}))^2}}}.
\end{eqnarray}
This can be inserted in Eq.\ref{inter-exciton-linearized}, which yields the gap equation:
\begin{eqnarray}
  \Delta(\vec k)  =
        {1 \over {4 } } \sum_{\vec k'}   V(|\vec k -\vec k'|)
       {{z(\vec k')} \over {
           \sqrt{|z(\vec k')|^2+(\delta\epsilon({\bf k}))^2}}},
           \label{eq:gap}
\end{eqnarray}
where $z(\vec k)$ is given by Eq.~\ref{zeta} in terms of $\Delta(\vec k)$
and the $w(\vec k)$ appearing in the SSH
Hamiltonian (Eq.~\ref{ssh-w}).
This equation can be solved self-consistently to find the
function $\Delta(\vec k)$. To this end a form for the
Coulomb interaction appropriate for a planar system is required.
We will use the same form to that
used in Ref.~\cite{PhysRevB.107.075105} to study the transition of a general 2D  band-insulator
to an excitonic-insulator.

\section{Results}
Since the diagonal term $\epsilon_{\nu}({\bf k})$ adds a topologically
trivial contribution, for the numerical
solution of the gap equation,  as an example, here, we
   consider the simpler case where we take $\epsilon_{\nu}({\bf k})=0$, and,
   in addition, we will also take $t'_d=t_d$.
This simplification gives rise to the model studied in Refs.~\cite{PhysRevB.100.104522,PhysRevB.106.054511} where ad hoc attractive {\it electron-electron}
interactions were added in order to investigate the possibility of Weyl superconductivity.

The gap $\Delta(k_x,k_y)$ was obtained by solving Eq.~\ref{eq:gap} iteratively until self-consistency was achieved. Fig.~\ref{fig5}(a) illustrates the
real part $\Delta_x$ and Fig.~\ref{fig5}(b) the imaginary part
$\Delta_y$ of the 
gap function
\begin{eqnarray}
  \Delta(\vec k)=\Delta_x(k_x,k_y) + i \Delta_y(k_x,k_y).
  \end{eqnarray}
Note that the
real part has a circular nodal line while the imaginary part has a
nodal line along the $k_x=0$ line. The two nodal lines intersect at the
two Weyl points,
which cause the two Weyl nodes seen in Fig.~\ref{fig5} which illustrates the
quasiparticle bands obtained from Eq.~\ref{quasiparticle} and
the self-consistent solution for the gap. The plus and minus signs
shown in Figs.~\ref{fig5} illustrate the location of the Weyl
nodes.

It is evident by inspecting Fig.~\ref{fig5},  which we verified numerically,
that near the
Weyl nodes, the real (imaginary) part of $\Delta(k_x,k_y)$ has the $p_y$ ($p_x$) symmetry, so the gap near the Weyl nodes has a $p_y + i p_x$ symmetry.
If we had rotated the geometry of our SSH lattice by $90^{\circ}$ the
symmetry would have been $p_x + i p_y$,  as in the case of
time-reversal symmetry-breaking pairing in unconventional superconductivity.

Fig.~\ref{fig7} illustrates the phase of the quasiparticle wavefunction (Eq.~\ref{qp-wave}),
defined by
\begin{eqnarray}
  \phi(k_x,k_y) \equiv \tan^{-1}\Bigl ({{z_y(k_x,k_y)} \over {z_x(k_x,k_y)}}  \Bigr ),\label{phase}
\end{eqnarray}
where
\begin{eqnarray}
  z(k_x,k_y)=z_x(k_x,k_y) + i z_y(k_x,k_y),
\end{eqnarray}
namely,
$z_{x}=\Delta_{x}(\vec k) - w_{x}(\vec k)$,
$z_{y}=\Delta_{y}(\vec k) + w_{y}(\vec k)$ are the real and imaginary parts of $z$. Thus, we can write the ratio
$z/|z|$ entering in Eq.~\ref{qp-wave} as $e^{i \phi}$ and
Fig.~\ref{fig7} illustrates $\phi$ inside the entire BZ.
Notice that the two points where the branch cut starts and ends, i.e.,
the Weyl points,  are shown
as + and - in the Figure. They are the beginning and the end of the branch cut via which you pass from the $-\pi/2$ (blue-colored)  to the $+\pi/2$ (red-colored) Riemann sheet. Notice as we travel around each of the Weyl nodes
the phase goes from the lower Riemann sheet characterized by a
$\mp\pi/2$ phase to the upper Riemann sheet of $\pm\pi/2$, registering a
phase change of $\pm\pi$.
 This  phase is directly related to
 the Berry-phase of particle-hole-paring wavefunction Berry-phase.
 We found, and this will become clear in the next paragraph,
that this approach, i.e., using the behavior
of the wavefunction relative phase to be a clearer way
to describe and demonstrate the topological
nature of the bands.

Taking $\epsilon_A({\bf k})=\epsilon_B({\bf k})=0$, the  $\hat {\mathrm{H}}_{\mathrm{BdG}}$ can be expresses as follows
\begin{eqnarray}
 \hat {\mathrm{H}}_{\mathrm{BdG}} &=&  \sum_{\vec k}  \Bigl (
           c^{\dagger}_{A}(\vec k),
             h_{B}(-\vec k) \Bigr )
       {\bf M}(\vec k)   \begin{pmatrix}
    c_A(\vec k) \\
      h^{\dagger}_B(-\vec k)  
  \end{pmatrix}, \label{ph} \\
  {\bf M}(\vec k) &=&  -{\vec B}({\vec k}) \cdot \vec \sigma \\
  {\vec B}(\vec k) &=& \begin{pmatrix}
    B_x(\vec k) \\
    B_y(\vec k) \\
    0 \\
  \end{pmatrix}, \\
  B_x(\vec k) &=& z_x(k_x,k_y), \hskip 0.2 in
  B_y(\vec k) = z_y(k_x,k_y).
\end{eqnarray}
$\vec B(\vec k)$ is a vector field in momentum space, which acts
as a pseudo-magnetic field through a Zeeman-like coupling, and it has
 two contributions one from the SSH Hamiltonian, i.e., from
the real and imaginary part of $w(\vec k)$ and a contribution
from the emergent collective behavior, i.e., from the
excitonic gap order parameter, i.e.,
\begin{eqnarray}
  {\vec B_{\mathrm{ind}}}(\vec k) &=& \begin{pmatrix}
    \Delta_x(\vec k) \\
    \Delta_y(\vec k) \\
    0 \\
  \end{pmatrix},
\end{eqnarray}
The lower (higher) energy band corresponds to the quantum pseudo-spin operator
$\vec \sigma$ having a projection parallel (anti-parallel) to the
the field $\vec B(\vec k)$.
In Fig.~\ref{fig8}(a) we plot the contribution to this field from
the order parameter $\Delta$. Near the two Weyl points (Fig.~\ref{fig8}(b-c))
$B(\vec k)$ is proportional to $\delta \vec k = \vec k - \vec k_{\pm}$,
where $\vec k_{\pm}$ is the location of the Weyl points.
Therefore, we obtain the same (pseudo-)spin-momentum correlations
to those which hold for the projection of the spin to momentum
in a continuum Weyl field-theory.
In all other points in the BZ, the pseudo-spin is
aligned with $\vec B$, and, therefore,
the average helicity of the ground state has the structure illustrated
in Fig.~\ref{fig8}.

We also calculated the Berry connection vector and has very similar structure
inside the BZ to that illustrated in Fig.~\ref{fig8}, and, thus, it would be
redundant to draw a very similar figure.
The Berry curvature in our case consists of just two
opposite-sign featureless $\delta$-function singularities at
the Weyl points and the net Berry-flux around each Weyl point is $\pm \pi$
as it becomes clear from Fig.~\ref{fig8}(b-c).
The net Berry-flux throughout the entirety of the BZ is zero because
the net topological charge is zero as the Weyl-nodes carry opposite
topological charges.

The edge-to-bulk correspondence present in a topology implies the
existence of edge states. So, we carried out a finite ribbon calculation where
the ribbon is infinite along the $k_y$-direction and finite along the
$x$-direction. The results obtained on a $L_x=400$ size ribbon and
periodic along
the $y$ direction are shown in Fig.~\ref{fig9}. We find two states localized
on the two opposite edges of the ribbon which form the Bogoliubov-deGennes
Fermi-arcs due to the fact that Berry-flux emanates from one bulk Weyl point
on the gap function and sinks in the other Weyl point.
We verified that these edge states are
Bogoliubov-Fermi arcs and not Majorana.
This could be related to the fact that unlike the case of superconducting
order, the particle-hole pairing in an excitonic ground state
does not lead to an off-diagonal long-range order in the two-body density
matrix.\cite{PhysRev.158.462,RevModPhys.40.755}.

\section{Conclusions}
\label{conclusions}
Weyl semimetals, as manifested at the single particle level, have been
experimentally discovered\cite{doi:10.1126/science.aaa9273,doi:10.1126/science.aaa9297,PhysRevX.5.031013} and have been
extensively discussed in the past 15 years\cite{Armitage2018},
including some of the role that Coulomb correlations can play\cite{PhysRevB.99.035123}.

Furthermore, excitonic insulators, i.e., where excitons form a BEC, have been proposed\cite{PhysRev.158.462,RevModPhys.40.755,PhysRevLett.74.1633,Eisenstein2004,science.aam6432,Butov_2004,RevModPhys.42.1}
and experimentally sought for quite for some time\cite{science.aam6432,Wang2019,PhysRevLett.99.146403,PhysRevB.90.155116,pnas.2010110118}.

The findings presented in the present paper consist 
an entirely new concept, a state of matter with unique character, where a
condensate of electron-hole pairs forms as a result of the interplay between topology and collective quantum many-body coherence.
We show that this excitonic condensation occurs with an order
parameter, the particle-hole pairing gap,  displaying unconventional 
topological features. Some of its properties can be summarized as follows:
i) The condensate is also characterized by electronic pseudo-spin
  chiral structure.
  ii) The gap is strongly anisotropic in momentum-space and both its real
  and imaginary part close at only special points of Weyl character.    
iii)   
  The  quasiparticles above this many-body ground state have well-defined
  pseudo-helicity and their electronic band-structure has  anisotropic
  Dirac-like  cones emanating from these singular Weyl points.
  The Berry-field associated with the phase of the collective excitonic
  ground-state wavefunction has a pseudo-spin-dipole-like texture. 
  iv)  There are chiral edge-states which are the Bogoliubov-deGennes
  Fermi-arc states. 

  \iffalse
  Note that this Weyl excitonic condensation does not occur in
  all cases of Weyl semimetals nor in all cases of interchain
  hopping in the 2D-SSH Hamiltonians.
   considered here.
  and found that, in some cases where the non-interacting 2D-SSH Hamiltonian
  is a generic Weyl semimetal, different in for inclusion of the long-range Coulomb
  electron-hole interaction leads to an excitonic insulator of trivial
  topological character.
  \fi
  We note that the results presented here  are for
    the case where
    $\delta t_d < \delta t/2$. If this condition is not satisfied, there are two pairs of
    Weyl nodes. The case of excitonic condensation
   and particle-hole pairing, when there are
   two pairs of Weyl points, is more complex and, perhaps more exotic, as we might have a more complex form of pairing
   or even quartetting (bi-excitonic condensation).
   The study of this case is beyond the scope of the present paper and
   we leave it for the future.
  Therefore, in the general 2D case, especially when the $A$ and $B$ sublattice atoms are not
  identical, the parameter space available for Weyl behavior is much more
  limited than in 3D\cite{Armitage2018,PhysRevB.99.035123}.
  Therefore, the chance to realize the Weyl excitonic condensation described
  here increases significantly in 3D materials.

Next, we wish to discuss how this state can be observed experimentally.
First, systems of ultra-cold atoms
in optical lattices have been shown to be simulators of idealized condensed matter systems with higher degree of flexibility and ``tunability'' to demonstrate
theoretical concepts and models. For example, the one-dimensional SSH model
has been simulated\cite{Atala2013} in such lattices of ultra cold-atoms.
Therefore, it is conceivable that an optical lattice demonstrating Weyl
behavior at the single particle level can be created\cite{PhysRevLett.114.225301}. The more tricky task, however, would be to generate a  ``synthetic'' attractive interaction
in the particle-hole channel in such systems.

In real condensed matter systems, obviously, there is no problem in finding
one where there is an attractive electron-hole interaction.
Finding one with Weyl points
at the single particle level is also generally not a difficult issue\cite{PhysRevB.99.035123}.
Crystals near a structural instability, such as
the one driven by an inversion symmetry breaking lattice distortion, such as
of Peierls or Jahn-Teller type, are good candidates to realize a more close form to the particular discussed here in the
2D-SSH  model.
One of the required key issues is that the behavior has to be that of
type-I Weyl with the Fermi level exactly at the Weyl points.

The main goal of this paper was to demonstrate that this quantum many-body coherent
state of matter with unusual topology is theoretically possible
in a well-defined, historically significant, and well-studied lattice model,
such as the coupled SSH chains. 
It can serve as a model to future work by guiding its experimental recognition in a scenario which
leads to its potential realization.

\section{Acknowledgments}

  This work was supported by the U.S. National Science Foundation under Grant No. NSF-DMR-2110814. 

\end{document}